%% file: main.tex
\pdfoutput=1 
\RequirePackage{build-config}
\newcommand{\formattype}{\formattypeACM} 
\newcommand{\acmType}{sigconf} 

\RequirePackage{ifthen}

\newcommand{\formattypeACM}{formattypeACM}
\newcommand{\formattypeIEEE}{formattypeIEEE}

\ifisdraft
  \ifthenelse{\equal{\formattype}{\formattypeIEEE}}{
    \documentclass[10pt,conference]{IEEEtran}
  } {
    \ifisarxiv
      \ifisanonymous
        \documentclass[\acmType,screen,nonacm,review,anonymous]{acmart}
        \settopmatter{printfolios=false,printccs=false,printacmref=false}
      \else
        \documentclass[\acmType,screen,nonacm,review]{acmart}
        \settopmatter{printfolios=false,printccs=false,printacmref=false}
      \fi
    \else
      \ifisanonymous
        \documentclass[\acmType,screen,review,anonymous]{acmart}
        \settopmatter{printfolios=false,printccs=true,printacmref=true}
      \else
        \documentclass[\acmType,screen,review]{acmart}
        \settopmatter{printfolios=false,printccs=true,printacmref=true}
      \fi
    \fi
  }
\else
  \ifthenelse{\equal{\formattype}{\formattypeIEEE}}{
    \documentclass[10pt,conference]{IEEEtran}
  } {
    \ifisarxiv
      \ifisanonymous
        \documentclass[\acmType,screen,nonacm,anonymous]{acmart}
        \settopmatter{printfolios=false,printccs=false,printacmref=false}
      \else
        \documentclass[\acmType,screen,nonacm]{acmart}
        \settopmatter{printfolios=false,printccs=false,printacmref=false}
      \fi
    \else
      \ifisanonymous
        \documentclass[\acmType,screen,anonymous]{acmart}
        \settopmatter{printfolios=false,printccs=true,printacmref=true}
      \else
        \documentclass[\acmType,screen]{acmart}
        \settopmatter{printfolios=false,printccs=true,printacmref=true}
      \fi
    \fi
  }
\fi

\ifthenelse{\equal{\formattype}{\formattypeIEEE}}{

} {
}

\newwrite\abstractoutput
\immediate\openout\abstractoutput=\jobname.abstract.output

\ifthenelse{\equal{\formattype}{\formattypeIEEE}}{
  \newcommand{\makeTitleAndAbstract}{
    \write\abstractoutput{\theAbstract}
    \maketitle
    \begin{abstract}
    \theAbstract
    \end{abstract}
  }
}{
  \newcommand{\makeTitleAndAbstract}{
    \write\abstractoutput{\theAbstract}
    \begin{abstract}
    \theAbstract
    \end{abstract}
    \maketitle
  }
}

\usepackage{paper-specific-macros}
\usepackage{package-config}

\usepackage{lipsum} 

\setcopyright{acmlicensed}
\acmPrice{15.00}
\acmDOI{10.1145/3611643.3613094}
\acmYear{2023}
\copyrightyear{2023}
\acmSubmissionID{fse23demo-p24-p}
\acmISBN{979-8-4007-0327-0/23/12}
\acmConference[ESEC/FSE '23]{Proceedings of the 31st ACM Joint European Software Engineering Conference and Symposium on the Foundations of Software Engineering}{December 3--9, 2023}{San Francisco, CA, USA}
\acmBooktitle{Proceedings of the 31st ACM Joint European Software Engineering Conference and Symposium on the Foundations of Software Engineering (ESEC/FSE '23), December 3--9, 2023, San Francisco, CA, USA}
\received{2023-05-11}
\received[accepted]{2023-07-20}



\begin{document}

\makeTitleAndAbstract

\section{Introduction}
\label{sec:intro}

Modern software development relies extensively on a complex network of
reusable open-source software components (packages). The largest~\cite{module-counts} repository
of packages is the NPM repository, which contains over three million packages,
and 35 million different versions of packages while serving tens of billions of downloads weekly.
Practically every JavaScript application depends on packages from the NPM repository.
Understanding the NPM ecosystem, including distribution properties, versioning,
and dependency relations is an important component for reasoning about 
JavaScript software development practices 
\cite{
https://ieeexplore.ieee.org/document/8468105,
https://ieeexplore.ieee.org/document/8812106},
security
\cite{
https://dl.acm.org/doi/10.1145/2001420.2001442,
https://dl.acm.org/doi/10.1145/1995376.1995398,
https://dl.acm.org/doi/10.1145/3460319.3464836}, 
program analysis
\cite{
https://dl.acm.org/doi/10.1145/1809028.1806598,
https://dl.acm.org/doi/10.1145/2491411.2491417,
https://dl.acm.org/doi/10.5555/2486788.2486887,
https://dl.acm.org/doi/10.1145/3428255}
and more.  

Existing package repository datasets, such as \texttt{libraries.io}~\cite{libraries.io} and DaSEA~\cite{DaSEA},
provide a wealth of information about dependency structure, author information, etc., even across multiple package ecosystems.
However, these datasets are typically limited in two ways:
\begin{inparaenum}
  \item only storing metadata and not code of packages; and
  \item not maintaining historic data when packages or versions of packages are unpublished or deleted from NPM.
\end{inparaenum}
Unfortunately, packages are often deleted from NPM.
Between July 12, 2022, and May 10, 2023, we have detected 335,325 versions of packages that have been deleted.
This loss of data is problematic for data availability and artifact reproducibility for research which may use package data
(regression testing~\cite{https://drops.dagstuhl.de/opus/volltexte/2018/9212}, static analysis~\cite{https://dl.acm.org/doi/10.5555/2486788.2486887}, training large-language models~\cite{https://arxiv.org/abs/2211.15533}, etc.),
and makes research areas that specifically examine deleted packages (such as malware analysis) 
nearly impossible without privileged access~\cite{amalfi}.

We present \theName{} as a platform to enable easier research on the NPM ecosystem. We believe that \theName{} offers two main benefits.
First, \theName{} continually collects and archives packages, and thus retains data (including package code\footnote{Note that \theName{} collects the code which is released by developers to NPM,
which may be different from the source code of a project's GitHub repository (\cref{sec:related}).})
which is later deleted from NPM.
Second, \theName{} scrapes and indexes multiple sources of data (developer-provided metadata, code, download metrics and security advisories),
allowing researchers to easily write analyses which touch many aspects of the NPM ecosystem.
The \theName{} dataset and source code is available
at \url{https://dependencies.science}, and we hope that it will be useful to the research community.

\section{Related Work}
\label{sec:related}

A variety of existing tools scrape data from software ecosystems.
Roughly, these can be divided into two areas: those that include metadata only, and those that store source code.

The \texttt{libraries.io}~\cite{libraries.io} website hosts a dataset of metadata spanning multiple package managers,
and is quite detailed. However, \citet{DaSEA} report that it is not well maintained and does not include metadata for all versions of packages,
and in response introduced the DaSEA dataset which is a cross-ecosystem dataset containing metadata for versions of packages.
Unfortunately, DaSEA does not include NPM (the largest and fastest growing repository~\cite{module-counts}).
In addition, neither \texttt{libraries.io} nor DaSEA store package code themselves.
Thus, to perform package code analysis one would have to download package code from
NPM, which is time and labor intensive, and is impossible for packages that have been deleted from NPM.

On the other hand, large-scale projects exist which archive not only metadata but also source code.
GHTorrent~\cite{gh-torrent} and World of Code~\cite{WoC} collect and archive source code from VCS hosting platforms (GitHub, etc.).
Unfortunately, GHTorrent appears to be unmaintained, and both focus on scraping VCS data rather than package manager repositories. Packages uploaded to NPM do not necessarily have an associated (public) VCS repository,
and code uploaded to NPM may in fact be different from source code in a VCS repository, so these are related but complementary sources of data.
The Software Heritage archive~\cite{software-heritage} collects the full source code of packages across
multiple software ecosystems, including NPM. However, the Software Heritage archive performs intermittent scraping,
similar to the Wayback Machine~\cite{wayback-machine}, so it is not able to download packages which are uploaded
and deleted in-between scrapes. In contrast, \theName{} receives updates from the NPM repository and downloads new
packages as they appear with low latency (\cref{sec:scraping-blob}).

Using software ecosystem datasets, 
researchers are able to examine many interesting research topics, such as 
technical lag~\cite{msr-resp-c1}, 
versioning~\cite{maven-semver-not-followed,dynamics-of-js-ecosystem}, 
micro packages~\cite{micro-packages},
static analysis~\cite{https://dl.acm.org/doi/10.5555/2486788.2486887}, 
malware analysis~\cite{weak-links-npm,amalfi,npm-security-threats,msr-resp-a1}, 
security vulnerability analysis~\cite{msr-resp-d1,msr-resp-b4} and more,
all of which need access to either metadata or package source code data.
In our prior work, we used NPM ecosystem data to evaluate our technique for optimal dependency solving~\cite{maxnpm}, 
and to understand how developers make use of semantic versioning and updates in practice~\cite{npmdata}.
The first version of \theName{} was born out of that work, and since then we have provided more built-in analyses,
added scraping of package download metrics, and continued to improve reliability. In this paper we discuss key design decisions of \theName{},
as well as challenges for sustainability of the system.

\section{Using \theName{}}
\label{sec:uses}

\input{figures/architecture2}

The \theName{} dataset is useful for answering research questions involving metadata and/or source code analysis, 
such as evaluation of static analysis tooling, 
detection of malware, or training code large-language models.
To illustrate how \theName{} could be used in such research,
we present a hypothetical example of vulnerability impact analysis~\cite{impact-1,impact-2},
which has the goal of identifying client libraries that may be impacted by a security vulnerability in a dependency,
and if the vulnerable code is in fact reachable from the client.
We can use \theName{} to perform the first half of that task: finding pairs of clients and dependencies, where
the dependency has a vulnerability. To do so, we will first work on building the set of dependencies,
and then match them with dependent clients. A variant of this example is available as a video demonstration.
\footnote{\url{https://youtu.be/OgLYThRJhdc?si=V6krLg3LzUvUeH7u}}

\subsection{Finding Packages with Vulnerabilities}

In addition to package metadata, \theName{} also scrapes security advisories from the GitHub Advisory Database.
We can find packages that have vulnerabilities by joining the table of packages (center of \cref{fig:db-layout})
with the table of vulnerabilities (top of \cref{fig:db-layout}):

\begin{minted}[
  % numbersep=5pt, 
  fontfamily=zi4,
  fontsize=\footnotesize]{sql}
  select ...
  from packages vuln_p 
  join vulnerabilities vuln on vuln_p.name = vuln.package_name
\end{minted}

However, in order to obtain a smaller, more focused dataset we may wish to only
select packages which also have a decent number of downloads. 
We may accomplish this by additionally joining scraped download metrics (bottom of \cref{fig:db-layout}) 
and keeping only those with over 1 million weekly downloads:

\begin{minted}[
  % numbersep=5pt, 
  fontfamily=zi4,
  fontsize=\footnotesize]{sql}
  ...
  join download_metrics m on m.package_id = vuln_p.id
  and (m.download_counts[array_upper(m.download_counts, 1)]).counter
    > 1000000
\end{minted}

\noindent Note that the code outlined here does not distinguish different versions of a vulnerable
package, even though typically vulnerabilities only affect some versions. A more complex
analysis that selects specific versions of packages that are vulnerable is possible
with \theName{}, and in fact already has a reusable implementation (\cref{sec:querying-metadata}).

\subsection{Determining Dependent Clients}

Now that we have a set of packages that contain security vulnerabilities,
we can find a corresponding set of dependent client package versions
by using a relation describing the dependencies of each version of each package:

\begin{minted}[
  % numbersep=5pt, 
  fontfamily=zi4,
  fontsize=\footnotesize,
  escapeinside=||]{sql}
  ...
  join metadata_analysis.version_direct_runtime_deps edge 
    on edge.depends_on_pkg = vuln_p.id
  join |versions| client on client.id = edge.v
\end{minted}

This dependency relation table is not part of the core data model of \theName{}
but is computed from the core tables with a provided analysis implementation.
As above, this simple query does not take into account the version of the vulnerable
package, so it may be that the clients depend on non-vulnerable versions. If needed,
this may be addressed by writing a more complex query on the dependency version
constraint data structure (\cref{sec:querying-metadata}).

Finally, if we aim to use dynamic analysis techniques to look at vulnerability impact
in the clients, we may wish to only select clients which have tests.
Since \theName{} stores in original JSON format all metadata it does not specifically extract,
we may use this to filter for tests:

\begin{minted}[
  % numbersep=5pt, 
  fontfamily=zi4,
  fontsize=\footnotesize]{sql}
  ...
  and client.extra_metadata->'scripts'->'test' is not null
\end{minted}

We have now completed finding our set of client and vulnerable library pairs we wish to
analyze, and can move on to retrieving package code for these pairs.

\subsection{Obtaining Package Code}

We now would like to obtain the code, say for the clients, to proceed with our vulnerability impact analysis.
One way would be to read the \texttt{client.tarball\_url} column from the query above and download each tarball. Unfortunately,
some of these URLs might return 404 errors because developers could have unpublished those versions due to depending on a vulnerability,
leading to obtaining a biased sample. 

A different approach would be to use the \theName{} code store (\cref{sec:scraping-blob}), 
which attempts to archive
tarballs before they are deleted. Doing so is easy, as all source code URLs map into the object store, which can then be read from
using \theName{}'s tooling. 
After obtaining datasets of lists of vulnerable packages and clients and associated source code, we are
now in a good position to explore exciting research techniques to determine vulnerability impact.

\section{Design and Implementation}
\label{sec:scraping}

When designing \theName{}, we had two primary design goals:
\begin{inparaenum}
  \item it should be a comprehensive and easy to use
  dataset for analysis of the NPM ecosystem; and
  \item it should be able to be sustainably run using our available hardware resources.
\end{inparaenum}
Specifically, we have available an academic 
Slurm-backed~\cite{slurm-website} HPC cluster with around \qty{25}{TB} of networked file storage, which we use for downloading and storing tarballs of package code.
The metadata portion of \theName{} is able to be run independently, on a single Linux VM with \qty{4}{CPUs} and \qty{128}{GB} of RAM, currently requiring
about \qty{700}{GB} of SSD storage. 


\subsection{Package Metadata}

While the most obvious contribution of \theName{} is in the scraping and storing of package code data,
we nevertheless designed the metadata scraping and analysis component of \theName{} to behave
well in the presence of package deletions, and to use a richer data model than prior work to
enable more complex analyses. 
The metadata of \theName{}
is stored in PostgreSQL~\cite{postgres-website}, which provides for a consistent
and easy to use data analysis platform while providing sufficient throughput
to index updates from NPM.

\subsubsection{Streaming and Parsing Updates}
NPM offers a changes streaming API~\cite{npm-changes-github}
which allows us to stream updates (package / version creation and deletion operations) in a JSON format
without needing to frequently crawl the NPM website.
Unfortunately, the raw JSON updates have two major problems:
\begin{inparaenum}
  \item the data is poorly structured with little validation, making basic data analysis difficult; and
  \item when a new version of a package is published, the corresponding change notification contains all previous versions in addition to the new version, which causes storage to grow quadratically with the number of updates if naively storing all change notifications.
\end{inparaenum}

To better parse and index the metadata, \theName{} first
validates and cleans update events as it receives them, and inserts the data into the relational database while de-duplicating repetitive data.
In addition, we carefully parse common metadata fields which are useful for data analysis into interpretable data structures, 
including version numbers, 
dependency version constraints, and GitHub repository data. 
Metadata fields which we don't specifically parse (e.g. author's names and emails) or fail to parse (e.g. invalid source code repository URIs) are retained in JSON format and available for querying.
Additionally, data fields are not deleted when a package (or version of a package) is deleted from NPM.
Instead, the entity is only marked with a deleted flag.

\subsubsection{Querying Package Metadata}
\label{sec:querying-metadata}

Three tables store the metadata obtained from the NPM changes API: 
\texttt{packages}, \texttt{versions} and \texttt{dependencies}, 
which collectively enumerate all versions of all packages, and the dependencies of each version. 
These tables make up the core data model of \theName{} (center of \cref{fig:db-layout}), 
and are typically the starting point of analyses.

Since \theName{} performs parsing on many metadata fields,
it is often possible to write SQL queries which directly interpret these fields. 
The most interesting example is dependency version constraints, which are parsed
from their string format into disjunctive
normal form (DNF) over the total ordering of version numbers.
For example, the version constraint \texttt{``12 || ~13.0.1''} would be parsed into
$
  (X \geq 12.0.0 \land X < 13.0.0) \lor (X \geq 13.0.1 \land X < 13.1.0)
$,
and finding a matching version then corresponds to finding a satisfying assignment 
for $X$ drawn from the set of versions of the dependency.
Matching versions can be computed in SQL by matching candidate versions to each ordering term, and then
aggregating conjunctions followed by aggregating disjunctions.

In contrast, prior work~\cite{libraries.io,DaSEA}
only provides version constraints as uninterpreted strings, and leaves it up to the user of the dataset
to interpret the constraints if desired.
Interpreting constraints correctly either requires substantial
work~\cite{maxnpm}, or forces the analysis pipeline to interoperate with a JavaScript package for interpreting
version constraints~\cite{npm-semver}.

Since writing queries that operate on these interpreted data structures is non-trivial,
\theName{} includes a small library of common analyses that users
may wish to use and build on. Some of these analyses include computing
updates between versions of packages\footnote{tricky because version ordering and temporal ordering need not agree, see our prior work for details on this analysis~\cite{npmdata}.},
resolutions of direct dependency version constraints,
transitive dependency graph computation,
and identifying versions of packages which contain a security vulnerability.

\subsection{Code Acquisition and Storage}
\label{sec:scraping-blob}

When \theName{} receives metadata updates that contain URLs to new package code tarballs, 
\theName{} enqueues a download job to then be handled by the code data downloading and storage subsystem running on our HPC cluster.

Storing package code data is challenging, 
due to both the scale (tens of millions of tarballs, 20+\,TB) 
and the need to handle sufficient concurrent writes.
We did not explore using existing distributed file systems such as Hadoop~\cite{hadoop-website} due to concern of Hadoop's scalability with regards
to storing many small files~\cite{hadoop-small-files} (our use case). In addition, we are unsure if Hadoop
can run correctly and efficiently on top of the networked file system at our disposal.

Instead, we store tarball data in a custom-built object storage system stored on the networked file system. 
A manager node controls access to the object storage, keeping track of byte offsets and coordinating locks for writing,
while individual worker nodes in the HPC cluster perform the networked disk I/O.
Download jobs are dequeued from the work queue (enqueued by the metadata subsystem) and assigned to worker nodes in the HPC cluster.

We have observed that the download latency (tarball published to NPM $\to$ downloaded by \theName{})
has a bimodal distribution, with one group of tarballs having quite low latency (about \qty{30}{s}),
and another group of tarballs having higher latency (\qty{1}{hour} -- \qty{1}{day}). The higher latency downloads appear correlated
with periods of higher load, and may be caused both by unavoidable latency in NPM sending change notifications,
and latency within \theName{} itself. Overall, we are able to download 98.8\% of tarballs within a latency of
\qty{24}{hours}, which is satisfactory for our purposes.

To allow data analysis jobs to read from the package code data (right side of \cref{fig:db-layout}), 
the manager node exposes a mapping of
tarball keys (derived from the \texttt{downloaded\_tarballs} table) to underlying file system location information (file name, byte offset, number of bytes). To read a package code tarball,
a worker node first queries the manager node for the location on disk, and then itself performs the read from the underlying (networked)
file system. Fortunately, our software abstracts over this separation, allowing for a simple \texttt{cp}-like command to read
a file out of the object store given a key. This system allows for large-scale concurrent reading from the object storage to perform code analysis.

\subsubsection{Tarball Size Distribution}

Currently the object storage system is about 24 TB in size and stores over 35 million tarballs.
However, the distribution of tarball sizes is extremely skewed (median = \qty{18.4}{KB}, mean = \qty{730}{KB}),
with the largest tarball being over \qty{500}{MB}. Based on this skewed distribution, one could consider
trading-off completeness for storage size. For instance, discarding all tarballs greater than \qty{16}{MB}
would cut the total size in half, while retaining 99.12\% of all tarballs.
While most of these oversize tarballs belong to obscure packages, sprinkled among them are popular packages, 
such as the NPM CLI (\qty{52}{MB}, 890+ million downloads) 
and \texttt{gherkin} (\qty{120}{MB}, 187+ million downloads).
In the future we plan to investigate better discarding strategies by incorporating both 
tarball size and download metrics, in order to keep the storage requirements for \theName{} sustainable.

\subsection{Scraping External Metadata}
\label{sec:other-scrapers}

Additionally, \theName{} scrapes two other sources of metadata: 
security advisory metadata from the GitHub Security Advisory Database (top of \cref{fig:db-layout}) 
and package download metrics (bottom of \cref{fig:db-layout}).

The security advisory metadata lists security advisories for NPM packages, which versions of packages are vulnerable,
and applicable CWEs. 
Our prior work used this security metadata in a prototype of \theName{} to analyze the relationships
between semantic versioning and security effects~\cite{npmdata}.
The package download metrics provide weekly time-series data on the number of downloads each package receives,
and are often useful for pre-filtering data prior to other analyses, such as to focus on the top $N$ downloaded packages.

Unlike the metadata of packages, for both security metadata and download metrics we do not
have a convenient changes API which can notify us of new data,
and in particular the scraping of package download metrics~\cite{npm-download-metrics-api} is challenging due to severe rate-limiting.
To scrape download metrics for all packages in a reasonable
amount of time, we must perform batch requests to the API, which unfortunately precludes 
scraping per-version download metrics. 
With this strategy, scraping download metrics for all packages takes about four days. 
We estimate that if we did not batch requests it would take about two weeks.
If rate limits were to increase, we could consider scraping per-version metrics in the future.

\subsection{Adaptability to Other Ecosystems}

In addition to NPM, many other package repositories are
important for software engineering, such as PyPI, APT and more. The general architecture of \theName{}
may be applied to create comprehensive datasets of other ecosystems, depending on available APIs.
The main requirement is a change notification API, 
which is crucial for enabling continual archiving of packages, 
and is central to the design of \theName{}.
Designing a unified data model for multiple ecosystems could be challenging while maintaining
easy querying and interpretable data structures, though prior work~\cite{libraries.io,DaSEA} has partially tackled this.
To have a unified interpretable format for version constraints, the DNF format of \theName{} may be able to used as a low-level
target to which high-level constraints of various diverse syntaxes are parsed into. 
Investigating generalizing \theName{} to allow for other ecosystems could be
an interesting direction for future work.

\section{Conclusion}
\label{sec:conclusion}

NPM is a quickly evolving and often unreliable archive of data,
as packages are deleted frequently. In this demonstration,
we have presented \theName{}, a scraper and dataset which
continually downloads and archives metadata and code from NPM packages.
We have further shown the utility of \theName{} for researchers working in the
area of program analysis or software ecosystem analysis. 
The code and dataset, featuring a complete account of the metadata and
code of all versions of all packages is available publicly at \url{https://dependencies.science}.

\section*{Acknowledgments}
This work was funded in part by NSF CCF-2102288, CCF-2100037 and NSF CNS-2100015.

{
  \footnotesize

  \ifthenelse{\equal{\formattype}{\formattypeIEEE}} {
    \bibliographystyle{IEEEtran}
  } {
    \bibliographystyle{ACM-Reference-Format}
  }

  \bibliography{bib/venues-short,bib/main}
}

\end{document}

%% file: figures/architecture2.tex
\tikzset{
  module/.style={%
      draw, rounded corners,
      minimum width=#1,
      minimum height=7mm,
      font=\sffamily
    },
  module/.default=2cm,
  >=LaTeX
}
\tikzset{database/.style={cylinder,aspect=0.5,draw,rotate=90,path picture={
          \draw (path picture bounding box.160) to[out=180,in=180] (path picture bounding
          box.20);
          \draw (path picture bounding box.200) to[out=180,in=180] (path picture bounding
          box.340);
        }}}
\makeatletter
\pgfdeclareshape{document}{
  \inheritsavedanchors[from=rectangle] 
  \inheritanchorborder[from=rectangle]
  \inheritanchor[from=rectangle]{center}
  \inheritanchor[from=rectangle]{north}
  \inheritanchor[from=rectangle]{south}
  \inheritanchor[from=rectangle]{west}
  \inheritanchor[from=rectangle]{east}
  \backgroundpath{
    \southwest \pgf@xa=\pgf@x \pgf@ya=\pgf@y
    \northeast \pgf@xb=\pgf@x \pgf@yb=\pgf@y
    \pgf@xc=\pgf@xb \advance\pgf@xc by-7.5pt 
    \pgf@yc=\pgf@yb \advance\pgf@yc by-7.5pt
    \pgfpathmoveto{\pgfpoint{\pgf@xa}{\pgf@ya}}
    \pgfpathlineto{\pgfpoint{\pgf@xa}{\pgf@yb}}
    \pgfpathlineto{\pgfpoint{\pgf@xc}{\pgf@yb}}
    \pgfpathlineto{\pgfpoint{\pgf@xb}{\pgf@yc}}
    \pgfpathlineto{\pgfpoint{\pgf@xb}{\pgf@ya}}
    \pgfpathclose
    \pgfpathmoveto{\pgfpoint{\pgf@xc}{\pgf@yb}}
    \pgfpathlineto{\pgfpoint{\pgf@xc}{\pgf@yc}}
    \pgfpathlineto{\pgfpoint{\pgf@xb}{\pgf@yc}}
    \pgfpathlineto{\pgfpoint{\pgf@xc}{\pgf@yc}}
  }
}
\makeatother

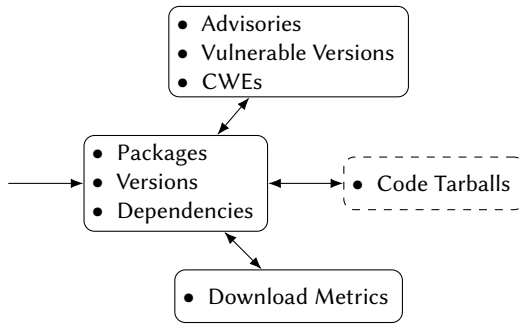
\begin{figure}
  \centering
    \begin{tikzpicture}
      \node[module, draw, align=center] (core) {
        \begin{varwidth}{\linewidth}\begin{itemize}[leftmargin=*]
            \item Packages
            \item Versions
            \item Dependencies
        \end{itemize}\end{varwidth}
      };

      \node[module, draw, align=center, above=of core, xshift=1.5cm, yshift=-0.5cm] (sec) {
        \begin{varwidth}{\linewidth}\begin{itemize}[leftmargin=*]
            \item Advisories
            \item Vulnerable Versions
            \item CWEs
        \end{itemize}\end{varwidth}
      };

      \node[module, draw, align=center, below=of core, xshift=1.5cm, yshift=0.5cm] (down) {
        \begin{varwidth}{\linewidth}\begin{itemize}[leftmargin=*]
            \item Download Metrics
        \end{itemize}\end{varwidth}
      };

      \node[module, dashed, align=center, right=of core] (code) {
        \begin{varwidth}{\linewidth}\begin{itemize}[leftmargin=*]
            \item Code Tarballs
        \end{itemize}\end{varwidth}
      };

      \node[left=of core] (entry) {};

      \draw[<->] (core)--(sec);
      \draw[<->] (core)--(down);
      \draw[<->] (core)--(code);
      \draw[->] (entry)--(core);
    \end{tikzpicture}
  \caption{
    Conceptual structure of the \theName{} database.
  }
  \label{fig:db-layout}
\end{figure}

%% file: main.bbl

\begin{thebibliography}{39}


\ifx \showCODEN    \undefined \def \showCODEN     #1{\unskip}     \fi
\ifx \showDOI      \undefined \def \showDOI       #1{#1}\fi
\ifx \showISBNx    \undefined \def \showISBNx     #1{\unskip}     \fi
\ifx \showISBNxiii \undefined \def \showISBNxiii  #1{\unskip}     \fi
\ifx \showISSN     \undefined \def \showISSN      #1{\unskip}     \fi
\ifx \showLCCN     \undefined \def \showLCCN      #1{\unskip}     \fi
\ifx \shownote     \undefined \def \shownote      #1{#1}          \fi
\ifx \showarticletitle \undefined \def \showarticletitle #1{#1}   \fi
\ifx \showURL      \undefined \def \showURL       {\relax}        \fi
\providecommand\bibfield[2]{#2}
\providecommand\bibinfo[2]{#2}
\providecommand\natexlab[1]{#1}
\providecommand\showeprint[2][]{arXiv:#2}

\bibitem[Archive(2023)]%
        {wayback-machine}
\bibfield{author}{\bibinfo{person}{Internet Archive}.}
  \bibinfo{year}{2023}\natexlab{}.
\newblock \bibinfo{title}{Wayback Machine}.
\newblock \bibinfo{howpublished}{\url{https://web.archive.org}. Accessed May 5
  2023}.
\newblock


\bibitem[Balint(2009)]%
        {hadoop-small-files}
\bibfield{author}{\bibinfo{person}{Szele Balint}.}
  \bibinfo{year}{2009}\natexlab{}.
\newblock \bibinfo{title}{The Small Files Problem}.
\newblock
  \bibinfo{howpublished}{\url{https://blog.cloudera.com/the-small-files-problem/}.
  Accessed Mar 13 2023}.
\newblock


\bibitem[Bandhakavi et~al\mbox{.}(2011)]%
        {https://dl.acm.org/doi/10.1145/1995376.1995398}
\bibfield{author}{\bibinfo{person}{Sruthi Bandhakavi}, \bibinfo{person}{Nandit
  Tiku}, \bibinfo{person}{Wyatt Pittman}, \bibinfo{person}{Samuel~T. King},
  \bibinfo{person}{P. Madhusudan}, {and} \bibinfo{person}{Marianne Winslett}.}
  \bibinfo{year}{2011}\natexlab{}.
\newblock \showarticletitle{Vetting Browser Extensions for Security
  Vulnerabilities with VEX}.
\newblock \bibinfo{journal}{\emph{Commun. ACM}} \bibinfo{volume}{54},
  \bibinfo{number}{9} (\bibinfo{date}{sep} \bibinfo{year}{2011}),
  \bibinfo{pages}{91–99}.
\newblock
\showISSN{0001-0782}
\urldef\tempurl%
\url{https://doi.org/10.1145/1995376.1995398}
\showDOI{\tempurl}


\bibitem[Buchkova et~al\mbox{.}(2022)]%
        {DaSEA}
\bibfield{author}{\bibinfo{person}{Petya Buchkova}, \bibinfo{person}{Joakim~Hey
  Hinnerskov}, \bibinfo{person}{Kasper Olsen}, {and}
  \bibinfo{person}{Rolf-Helge Pfeiffer}.} \bibinfo{year}{2022}\natexlab{}.
\newblock \showarticletitle{DaSEA: A Dataset for Software Ecosystem Analysis}.
  In \bibinfo{booktitle}{\emph{Proceedings of the 19th International Conference
  on Mining Software Repositories}} (Pittsburgh, Pennsylvania)
  \emph{(\bibinfo{series}{MSR '22})}. \bibinfo{publisher}{Association for
  Computing Machinery}, \bibinfo{address}{New York, NY, USA},
  \bibinfo{pages}{388–392}.
\newblock
\showISBNx{9781450393034}
\urldef\tempurl%
\url{https://doi.org/10.1145/3524842.3528004}
\showDOI{\tempurl}


\bibitem[Chinthanet et~al\mbox{.}(2021)]%
        {msr-resp-b4}
\bibfield{author}{\bibinfo{person}{Bodin Chinthanet},
  \bibinfo{person}{Raula~Gaikovina Kula}, \bibinfo{person}{Shane McIntosh},
  \bibinfo{person}{Takashi Ishio}, \bibinfo{person}{Akinori Ihara}, {and}
  \bibinfo{person}{Kenichi Matsumoto}.} \bibinfo{year}{2021}\natexlab{}.
\newblock \showarticletitle{Lags in the release, adoption, and propagation of
  npm vulnerability fixes}.
\newblock \bibinfo{journal}{\emph{Empirical Software Engineering}}
  \bibinfo{volume}{26}, \bibinfo{number}{3} (\bibinfo{date}{30 Mar}
  \bibinfo{year}{2021}), \bibinfo{pages}{47}.
\newblock
\showISSN{1573-7616}
\urldef\tempurl%
\url{https://doi.org/10.1007/s10664-021-09951-x}
\showDOI{\tempurl}


\bibitem[DeBill(2023)]%
        {module-counts}
\bibfield{author}{\bibinfo{person}{Erik DeBill}.}
  \bibinfo{year}{2023}\natexlab{}.
\newblock \bibinfo{title}{Modulecounts}.
\newblock \bibinfo{howpublished}{\url{http://www.modulecounts.com}. Accessed
  May 5 2023}.
\newblock


\bibitem[Decan et~al\mbox{.}(2018)]%
        {msr-resp-d1}
\bibfield{author}{\bibinfo{person}{Alexandre Decan}, \bibinfo{person}{Tom
  Mens}, {and} \bibinfo{person}{Eleni Constantinou}.}
  \bibinfo{year}{2018}\natexlab{}.
\newblock \showarticletitle{On the Impact of Security Vulnerabilities in the
  Npm Package Dependency Network}. In \bibinfo{booktitle}{\emph{Proceedings of
  the 15th International Conference on Mining Software Repositories}}
  (Gothenburg, Sweden) \emph{(\bibinfo{series}{MSR '18})}.
  \bibinfo{publisher}{Association for Computing Machinery},
  \bibinfo{address}{New York, NY, USA}, \bibinfo{pages}{181–191}.
\newblock
\showISBNx{9781450357166}
\urldef\tempurl%
\url{https://doi.org/10.1145/3196398.3196401}
\showDOI{\tempurl}


\bibitem[Feldthaus et~al\mbox{.}(2013)]%
        {https://dl.acm.org/doi/10.5555/2486788.2486887}
\bibfield{author}{\bibinfo{person}{Asger Feldthaus}, \bibinfo{person}{Max
  Schäfer}, \bibinfo{person}{Manu Sridharan}, \bibinfo{person}{Julian Dolby},
  {and} \bibinfo{person}{Frank Tip}.} \bibinfo{year}{2013}\natexlab{}.
\newblock \showarticletitle{Efficient construction of approximate call graphs
  for JavaScript IDE services}. In \bibinfo{booktitle}{\emph{2013 35th
  International Conference on Software Engineering (ICSE)}}.
  \bibinfo{pages}{752--761}.
\newblock
\urldef\tempurl%
\url{https://doi.org/10.1109/ICSE.2013.6606621}
\showDOI{\tempurl}


\bibitem[Foundation(2023)]%
        {hadoop-website}
\bibfield{author}{\bibinfo{person}{The Apache~Software Foundation}.}
  \bibinfo{year}{2023}\natexlab{}.
\newblock \bibinfo{title}{Apache Hadoop}.
\newblock \bibinfo{howpublished}{\url{https://hadoop.apache.org}. Accessed Mar
  13 2023}.
\newblock


\bibitem[Gonzalez-Barahona et~al\mbox{.}(2017)]%
        {msr-resp-c1}
\bibfield{author}{\bibinfo{person}{Jesus~M. Gonzalez-Barahona},
  \bibinfo{person}{Paul Sherwood}, \bibinfo{person}{Gregorio Robles}, {and}
  \bibinfo{person}{Daniel Izquierdo}.} \bibinfo{year}{2017}\natexlab{}.
\newblock \showarticletitle{Technical Lag in Software Compilations: Measuring
  How Outdated a Software Deployment Is}. In \bibinfo{booktitle}{\emph{Open
  Source Systems: Towards Robust Practices}},
  \bibfield{editor}{\bibinfo{person}{Federico Balaguer},
  \bibinfo{person}{Roberto Di~Cosmo}, \bibinfo{person}{Alejandra Garrido},
  \bibinfo{person}{Fabio Kon}, \bibinfo{person}{Gregorio Robles}, {and}
  \bibinfo{person}{Stefano Zacchiroli}} (Eds.). \bibinfo{publisher}{Springer
  International Publishing}, \bibinfo{address}{Cham},
  \bibinfo{pages}{182--192}.
\newblock
\showISBNx{978-3-319-57735-7}
\urldef\tempurl%
\url{https://doi.org/10.1007/978-3-319-57735-7_17}
\showDOI{\tempurl}


\bibitem[Gousios and Spinellis(2012)]%
        {gh-torrent}
\bibfield{author}{\bibinfo{person}{Georgios Gousios} {and}
  \bibinfo{person}{Diomidis Spinellis}.} \bibinfo{year}{2012}\natexlab{}.
\newblock \showarticletitle{GHTorrent: Github's data from a firehose}. In
  \bibinfo{booktitle}{\emph{2012 9th IEEE Working Conference on Mining Software
  Repositories (MSR)}}. \bibinfo{pages}{12--21}.
\newblock
\urldef\tempurl%
\url{https://doi.org/10.1109/MSR.2012.6224294}
\showDOI{\tempurl}


\bibitem[Guarnieri et~al\mbox{.}(2011)]%
        {https://dl.acm.org/doi/10.1145/2001420.2001442}
\bibfield{author}{\bibinfo{person}{Salvatore Guarnieri}, \bibinfo{person}{Marco
  Pistoia}, \bibinfo{person}{Omer Tripp}, \bibinfo{person}{Julian Dolby},
  \bibinfo{person}{Stephen Teilhet}, {and} \bibinfo{person}{Ryan Berg}.}
  \bibinfo{year}{2011}\natexlab{}.
\newblock \showarticletitle{Saving the World Wide Web from Vulnerable
  JavaScript}. In \bibinfo{booktitle}{\emph{Proceedings of the 2011
  International Symposium on Software Testing and Analysis}} (Toronto, Ontario,
  Canada) \emph{(\bibinfo{series}{ISSTA '11})}. \bibinfo{publisher}{Association
  for Computing Machinery}, \bibinfo{address}{New York, NY, USA},
  \bibinfo{pages}{177–187}.
\newblock
\showISBNx{9781450305624}
\urldef\tempurl%
\url{https://doi.org/10.1145/2001420.2001442}
\showDOI{\tempurl}


\bibitem[Heritage(2023)]%
        {software-heritage}
\bibfield{author}{\bibinfo{person}{Software Heritage}.}
  \bibinfo{year}{2023}\natexlab{}.
\newblock \bibinfo{title}{Software Heritage}.
\newblock \bibinfo{howpublished}{\url{https://www.softwareheritage.org}.
  Accessed May 5 2023}.
\newblock


\bibitem[Kavaler et~al\mbox{.}(2019)]%
        {https://ieeexplore.ieee.org/document/8812106}
\bibfield{author}{\bibinfo{person}{David Kavaler}, \bibinfo{person}{Asher
  Trockman}, \bibinfo{person}{Bogdan Vasilescu}, {and}
  \bibinfo{person}{Vladimir Filkov}.} \bibinfo{year}{2019}\natexlab{}.
\newblock \showarticletitle{Tool Choice Matters: JavaScript Quality Assurance
  Tools and Usage Outcomes in GitHub Projects}. In
  \bibinfo{booktitle}{\emph{2019 IEEE/ACM 41st International Conference on
  Software Engineering (ICSE)}}. \bibinfo{pages}{476--487}.
\newblock
\urldef\tempurl%
\url{https://doi.org/10.1109/ICSE.2019.00060}
\showDOI{\tempurl}


\bibitem[Kocetkov et~al\mbox{.}(2022)]%
        {https://arxiv.org/abs/2211.15533}
\bibfield{author}{\bibinfo{person}{Denis Kocetkov}, \bibinfo{person}{Raymond
  Li}, \bibinfo{person}{Loubna~Ben Allal}, \bibinfo{person}{Jia Li},
  \bibinfo{person}{Chenghao Mou}, \bibinfo{person}{Carlos~Muñoz Ferrandis},
  \bibinfo{person}{Yacine Jernite}, \bibinfo{person}{Margaret Mitchell},
  \bibinfo{person}{Sean Hughes}, \bibinfo{person}{Thomas Wolf},
  \bibinfo{person}{Dzmitry Bahdanau}, \bibinfo{person}{Leandro von Werra},
  {and} \bibinfo{person}{Harm de Vries}.} \bibinfo{year}{2022}\natexlab{}.
\newblock \bibinfo{title}{The Stack: 3 TB of permissively licensed source
  code}.
\newblock
\newblock
\urldef\tempurl%
\url{https://doi.org/10.48550/arXiv.2211.15533}
\showDOI{\tempurl}
\showeprint[arxiv]{2211.15533}~[cs.CL]


\bibitem[Kula et~al\mbox{.}(2017)]%
        {micro-packages}
\bibfield{author}{\bibinfo{person}{Raula~Gaikovina Kula}, \bibinfo{person}{Ali
  Ouni}, \bibinfo{person}{Daniel~M. German}, {and} \bibinfo{person}{Katsuro
  Inoue}.} \bibinfo{year}{2017}\natexlab{}.
\newblock \bibinfo{title}{On the Impact of Micro-Packages: An Empirical Study
  of the npm JavaScript Ecosystem}.
\newblock
\newblock
\urldef\tempurl%
\url{https://doi.org/10.48550/ARXIV.1709.04638}
\showDOI{\tempurl}


\bibitem[Ma et~al\mbox{.}(2021)]%
        {WoC}
\bibfield{author}{\bibinfo{person}{Yuxing Ma}, \bibinfo{person}{Tapajit Dey},
  \bibinfo{person}{Chris Bogart}, \bibinfo{person}{Sadika Amreen},
  \bibinfo{person}{Marat Valiev}, \bibinfo{person}{Adam Tutko},
  \bibinfo{person}{David Kennard}, \bibinfo{person}{Russell Zaretzki}, {and}
  \bibinfo{person}{Audris Mockus}.} \bibinfo{year}{2021}\natexlab{}.
\newblock \showarticletitle{World of Code: Enabling a Research Workflow for
  Mining and Analyzing the Universe of Open Source VCS Data}.
\newblock \bibinfo{journal}{\emph{Empirical Softw. Engg.}}
  \bibinfo{volume}{26}, \bibinfo{number}{2} (\bibinfo{date}{mar}
  \bibinfo{year}{2021}), \bibinfo{numpages}{42}~pages.
\newblock
\showISSN{1382-3256}
\urldef\tempurl%
\url{https://doi.org/10.1007/s10664-020-09905-9}
\showDOI{\tempurl}


\bibitem[Madsen et~al\mbox{.}(2013)]%
        {https://dl.acm.org/doi/10.1145/2491411.2491417}
\bibfield{author}{\bibinfo{person}{Magnus Madsen}, \bibinfo{person}{Benjamin
  Livshits}, {and} \bibinfo{person}{Michael Fanning}.}
  \bibinfo{year}{2013}\natexlab{}.
\newblock \showarticletitle{Practical Static Analysis of JavaScript
  Applications in the Presence of Frameworks and Libraries}. In
  \bibinfo{booktitle}{\emph{Proceedings of the 2013 9th Joint Meeting on
  Foundations of Software Engineering}} (Saint Petersburg, Russia)
  \emph{(\bibinfo{series}{ESEC/FSE 2013})}. \bibinfo{publisher}{Association for
  Computing Machinery}, \bibinfo{address}{New York, NY, USA},
  \bibinfo{pages}{499–509}.
\newblock
\showISBNx{9781450322379}
\urldef\tempurl%
\url{https://doi.org/10.1145/2491411.2491417}
\showDOI{\tempurl}


\bibitem[Mezzetti et~al\mbox{.}(2018)]%
        {https://drops.dagstuhl.de/opus/volltexte/2018/9212}
\bibfield{author}{\bibinfo{person}{Gianluca Mezzetti}, \bibinfo{person}{Anders
  M\o{}ller}, {and} \bibinfo{person}{Martin~Toldam Torp}.}
  \bibinfo{year}{2018}\natexlab{}.
\newblock \showarticletitle{{Type Regression Testing to Detect Breaking Changes
  in Node.js Libraries}}.
\newblock   \bibinfo{volume}{109} (\bibinfo{year}{2018}),
  \bibinfo{pages}{7:1--7:24}.
\newblock
\showISBNx{978-3-95977-079-8}
\showISSN{1868-8969}
\urldef\tempurl%
\url{https://doi.org/10.4230/LIPIcs.ECOOP.2018.7}
\showDOI{\tempurl}


\bibitem[M\o{}ller et~al\mbox{.}(2020)]%
        {https://dl.acm.org/doi/10.1145/3428255}
\bibfield{author}{\bibinfo{person}{Anders M\o{}ller},
  \bibinfo{person}{Benjamin~Barslev Nielsen}, {and}
  \bibinfo{person}{Martin~Toldam Torp}.} \bibinfo{year}{2020}\natexlab{}.
\newblock \showarticletitle{Detecting Locations in JavaScript Programs Affected
  by Breaking Library Changes}.
\newblock \bibinfo{journal}{\emph{Proc. ACM Program. Lang.}}
  \bibinfo{volume}{4}, \bibinfo{number}{OOPSLA}, Article
  \bibinfo{articleno}{187} (\bibinfo{date}{nov} \bibinfo{year}{2020}),
  \bibinfo{numpages}{25}~pages.
\newblock
\urldef\tempurl%
\url{https://doi.org/10.1145/3428255}
\showDOI{\tempurl}


\bibitem[Nielsen et~al\mbox{.}(2021)]%
        {https://dl.acm.org/doi/10.1145/3460319.3464836}
\bibfield{author}{\bibinfo{person}{Benjamin~Barslev Nielsen},
  \bibinfo{person}{Martin~Toldam Torp}, {and} \bibinfo{person}{Anders
  M\o{}ller}.} \bibinfo{year}{2021}\natexlab{}.
\newblock \showarticletitle{Modular Call Graph Construction for Security
  Scanning of Node.Js Applications}. In \bibinfo{booktitle}{\emph{Proceedings
  of the 30th ACM SIGSOFT International Symposium on Software Testing and
  Analysis}} (Virtual, Denmark) \emph{(\bibinfo{series}{ISSTA 2021})}.
  \bibinfo{publisher}{Association for Computing Machinery},
  \bibinfo{address}{New York, NY, USA}, \bibinfo{pages}{29–41}.
\newblock
\showISBNx{9781450384599}
\urldef\tempurl%
\url{https://doi.org/10.1145/3460319.3464836}
\showDOI{\tempurl}


\bibitem[{NPM}(2022)]%
        {npm-semver}
\bibfield{author}{\bibinfo{person}{{NPM}}.} \bibinfo{year}{2022}\natexlab{}.
\newblock \bibinfo{title}{semver(1) -- The semantic versioner for npm}.
\newblock \bibinfo{howpublished}{\url{https://github.com/npm/node-semver}}.
\newblock


\bibitem[NPM and Contributors(2022)]%
        {npm-download-metrics-api}
\bibfield{author}{\bibinfo{person}{NPM} {and} \bibinfo{person}{Contributors}.}
  \bibinfo{year}{2022}\natexlab{}.
\newblock \bibinfo{title}{package download counts}.
\newblock
  \bibinfo{howpublished}{\url{https://github.com/npm/registry/blob/1c794110badd54b9d9fb08e7489746b6089c6648/docs/download-counts.md}.
  Accessed Aug 19 2023}.
\newblock


\bibitem[NPM and Contributors(2023)]%
        {npm-changes-github}
\bibfield{author}{\bibinfo{person}{NPM} {and} \bibinfo{person}{Contributors}.}
  \bibinfo{year}{2023}\natexlab{}.
\newblock \bibinfo{title}{registry-follower-tutorial}.
\newblock
  \bibinfo{howpublished}{\url{https://github.com/npm/registry-follower-tutorial}.
  Accessed Mar 12 2023}.
\newblock


\bibitem[Ohm et~al\mbox{.}(2022)]%
        {msr-resp-a1}
\bibfield{author}{\bibinfo{person}{Marc Ohm}, \bibinfo{person}{Felix Boes},
  \bibinfo{person}{Christian Bungartz}, {and} \bibinfo{person}{Michael Meier}.}
  \bibinfo{year}{2022}\natexlab{}.
\newblock \showarticletitle{On the Feasibility of Supervised Machine Learning
  for the Detection of Malicious Software Packages}. In
  \bibinfo{booktitle}{\emph{Proceedings of the 17th International Conference on
  Availability, Reliability and Security}} (Vienna, Austria)
  \emph{(\bibinfo{series}{ARES '22})}. \bibinfo{publisher}{Association for
  Computing Machinery}, \bibinfo{address}{New York, NY, USA}, Article
  \bibinfo{articleno}{127}, \bibinfo{numpages}{10}~pages.
\newblock
\showISBNx{9781450396707}
\urldef\tempurl%
\url{https://doi.org/10.1145/3538969.3544415}
\showDOI{\tempurl}


\bibitem[Pinckney et~al\mbox{.}(2023a)]%
        {npmdata}
\bibfield{author}{\bibinfo{person}{D. Pinckney}, \bibinfo{person}{F. Cassano},
  \bibinfo{person}{A. Guha}, {and} \bibinfo{person}{J. Bell}.}
  \bibinfo{year}{2023}\natexlab{a}.
\newblock \showarticletitle{A Large Scale Analysis of Semantic Versioning in
  NPM}. In \bibinfo{booktitle}{\emph{2023 IEEE/ACM 20th International
  Conference on Mining Software Repositories (MSR)}}. \bibinfo{publisher}{IEEE
  Computer Society}, \bibinfo{address}{Los Alamitos, CA, USA},
  \bibinfo{pages}{485--497}.
\newblock
\urldef\tempurl%
\url{https://doi.org/10.1109/MSR59073.2023.00073}
\showDOI{\tempurl}


\bibitem[Pinckney et~al\mbox{.}(2023b)]%
        {maxnpm}
\bibfield{author}{\bibinfo{person}{Donald Pinckney}, \bibinfo{person}{Federico
  Cassano}, \bibinfo{person}{Arjun Guha}, \bibinfo{person}{Jonathan Bell},
  \bibinfo{person}{Massimiliano Culpo}, {and} \bibinfo{person}{Todd Gamblin}.}
  \bibinfo{year}{2023}\natexlab{b}.
\newblock \showarticletitle{Flexible and Optimal Dependency Management via
  Max-SMT}. In \bibinfo{booktitle}{\emph{Proceedings of the 45th International
  Conference on Software Engineering}} (Melbourne, Victoria, Australia)
  \emph{(\bibinfo{series}{ICSE '23})}. \bibinfo{publisher}{IEEE Press},
  \bibinfo{pages}{1418–1429}.
\newblock
\showISBNx{9781665457019}
\urldef\tempurl%
\url{https://doi.org/10.1109/ICSE48619.2023.00124}
\showDOI{\tempurl}


\bibitem[Ponta et~al\mbox{.}(2020)]%
        {impact-1}
\bibfield{author}{\bibinfo{person}{Serena~Elisa Ponta}, \bibinfo{person}{Henrik
  Plate}, {and} \bibinfo{person}{Antonino Sabetta}.}
  \bibinfo{year}{2020}\natexlab{}.
\newblock \showarticletitle{Detection, assessment and mitigation of
  vulnerabilities in open source dependencies}.
\newblock \bibinfo{journal}{\emph{Empirical Software Engineering}}
  \bibinfo{volume}{25}, \bibinfo{number}{5} (\bibinfo{date}{01 Sep}
  \bibinfo{year}{2020}), \bibinfo{pages}{3175--3215}.
\newblock
\showISSN{1573-7616}
\urldef\tempurl%
\url{https://doi.org/10.1007/s10664-020-09830-x}
\showDOI{\tempurl}


\bibitem[Raemaekers et~al\mbox{.}(2014)]%
        {maven-semver-not-followed}
\bibfield{author}{\bibinfo{person}{S. Raemaekers}, \bibinfo{person}{A. van
  Deursen}, {and} \bibinfo{person}{J. Visser}.}
  \bibinfo{year}{2014}\natexlab{}.
\newblock \showarticletitle{Semantic Versioning versus Breaking Changes: A
  Study of the Maven Repository}. In \bibinfo{booktitle}{\emph{2014 IEEE 14th
  International Working Conference on Source Code Analysis and Manipulation
  (SCAM)}}. \bibinfo{publisher}{IEEE Computer Society}, \bibinfo{address}{Los
  Alamitos, CA, USA}, \bibinfo{pages}{215--224}.
\newblock
\urldef\tempurl%
\url{https://doi.org/10.1109/SCAM.2014.30}
\showDOI{\tempurl}


\bibitem[Richards et~al\mbox{.}(2010)]%
        {https://dl.acm.org/doi/10.1145/1809028.1806598}
\bibfield{author}{\bibinfo{person}{Gregor Richards}, \bibinfo{person}{Sylvain
  Lebresne}, \bibinfo{person}{Brian Burg}, {and} \bibinfo{person}{Jan Vitek}.}
  \bibinfo{year}{2010}\natexlab{}.
\newblock \showarticletitle{An Analysis of the Dynamic Behavior of JavaScript
  Programs}. In \bibinfo{booktitle}{\emph{Proceedings of the 31st ACM SIGPLAN
  Conference on Programming Language Design and Implementation}} (Toronto,
  Ontario, Canada) \emph{(\bibinfo{series}{PLDI '10})}.
  \bibinfo{publisher}{Association for Computing Machinery},
  \bibinfo{address}{New York, NY, USA}, \bibinfo{pages}{1–12}.
\newblock
\showISBNx{9781450300193}
\urldef\tempurl%
\url{https://doi.org/10.1145/1806596.1806598}
\showDOI{\tempurl}


\bibitem[SchedMD and Contributors(2023)]%
        {slurm-website}
\bibfield{author}{\bibinfo{person}{SchedMD} {and}
  \bibinfo{person}{Contributors}.} \bibinfo{year}{2023}\natexlab{}.
\newblock \bibinfo{title}{Slurm Workload Manager -- Documentation}.
\newblock \bibinfo{howpublished}{\url{https://slurm.schedmd.com}. Accessed Mar
  12 2023}.
\newblock


\bibitem[Sejfia and Sch\"{a}fer(2022)]%
        {amalfi}
\bibfield{author}{\bibinfo{person}{Adriana Sejfia} {and} \bibinfo{person}{Max
  Sch\"{a}fer}.} \bibinfo{year}{2022}\natexlab{}.
\newblock \showarticletitle{Practical Automated Detection of Malicious Npm
  Packages}. In \bibinfo{booktitle}{\emph{Proceedings of the 44th International
  Conference on Software Engineering}} (Pittsburgh, Pennsylvania)
  \emph{(\bibinfo{series}{ICSE '22})}. \bibinfo{publisher}{Association for
  Computing Machinery}, \bibinfo{pages}{1681–1692}.
\newblock
\showISBNx{9781450392211}
\urldef\tempurl%
\url{https://doi.org/10.1145/3510003.3510104}
\showDOI{\tempurl}


\bibitem[Staicu and Pradel(2018)]%
        {impact-2}
\bibfield{author}{\bibinfo{person}{Cristian-Alexandru Staicu} {and}
  \bibinfo{person}{Michael Pradel}.} \bibinfo{year}{2018}\natexlab{}.
\newblock \showarticletitle{Freezing the Web: A Study of ReDoS Vulnerabilities
  in Javascript-Based Web Servers}. In \bibinfo{booktitle}{\emph{Proceedings of
  the 27th USENIX Conference on Security Symposium}} (Baltimore, MD, USA)
  \emph{(\bibinfo{series}{SEC'18})}. \bibinfo{publisher}{USENIX Association},
  \bibinfo{address}{USA}, \bibinfo{pages}{361–376}.
\newblock
\showISBNx{9781931971461}


\bibitem[{The PostgreSQL Global Development Group}(2023)]%
        {postgres-website}
\bibfield{author}{\bibinfo{person}{{The PostgreSQL Global Development Group}}.}
  \bibinfo{year}{2023}\natexlab{}.
\newblock \bibinfo{title}{PostgreSQL: The World's Most Advanced Open Source
  Relational Database}.
\newblock \bibinfo{howpublished}{\url{https://www.postgresql.org}. Accessed Mar
  12 2023}.
\newblock


\bibitem[Tidelift(2023)]%
        {libraries.io}
\bibfield{author}{\bibinfo{person}{Inc Tidelift}.}
  \bibinfo{year}{2023}\natexlab{}.
\newblock \bibinfo{title}{Libraries.io -- The Open Source Discovery Service}.
\newblock \bibinfo{howpublished}{\url{https://libraries.io}. Accessed May 5
  2023}.
\newblock


\bibitem[Tómasdóttir et~al\mbox{.}(2020)]%
        {https://ieeexplore.ieee.org/document/8468105}
\bibfield{author}{\bibinfo{person}{Kristín~Fjóla Tómasdóttir},
  \bibinfo{person}{Maurício Aniche}, {and} \bibinfo{person}{Arie
  Van~Deursen}.} \bibinfo{year}{2020}\natexlab{}.
\newblock \showarticletitle{The Adoption of JavaScript Linters in Practice: A
  Case Study on ESLint}.
\newblock \bibinfo{journal}{\emph{IEEE Transactions on Software Engineering}}
  \bibinfo{volume}{46}, \bibinfo{number}{8} (\bibinfo{year}{2020}),
  \bibinfo{pages}{863--891}.
\newblock
\urldef\tempurl%
\url{https://doi.org/10.1109/TSE.2018.2871058}
\showDOI{\tempurl}


\bibitem[Wittern et~al\mbox{.}(2016)]%
        {dynamics-of-js-ecosystem}
\bibfield{author}{\bibinfo{person}{Erik Wittern}, \bibinfo{person}{Philippe
  Suter}, {and} \bibinfo{person}{Shriram Rajagopalan}.}
  \bibinfo{year}{2016}\natexlab{}.
\newblock \showarticletitle{A Look at the Dynamics of the JavaScript Package
  Ecosystem}. In \bibinfo{booktitle}{\emph{Proceedings of the 13th
  International Conference on Mining Software Repositories}} (Austin, Texas)
  \emph{(\bibinfo{series}{MSR '16})}. \bibinfo{publisher}{Association for
  Computing Machinery}, \bibinfo{address}{New York, NY, USA},
  \bibinfo{pages}{351–361}.
\newblock
\showISBNx{9781450341868}
\urldef\tempurl%
\url{https://doi.org/10.1145/2901739.2901743}
\showDOI{\tempurl}


\bibitem[Zahan et~al\mbox{.}(2022)]%
        {weak-links-npm}
\bibfield{author}{\bibinfo{person}{Nusrat Zahan}, \bibinfo{person}{Thomas
  Zimmermann}, \bibinfo{person}{Patrice Godefroid}, \bibinfo{person}{Brendan
  Murphy}, \bibinfo{person}{Chandra Maddila}, {and} \bibinfo{person}{Laurie
  Williams}.} \bibinfo{year}{2022}\natexlab{}.
\newblock \showarticletitle{What Are Weak Links in the Npm Supply Chain?}. In
  \bibinfo{booktitle}{\emph{Proceedings of the 44th International Conference on
  Software Engineering: Software Engineering in Practice}} (Pittsburgh,
  Pennsylvania) \emph{(\bibinfo{series}{ICSE-SEIP '22})}.
  \bibinfo{publisher}{Association for Computing Machinery},
  \bibinfo{address}{New York, NY, USA}, \bibinfo{pages}{331–340}.
\newblock
\showISBNx{9781450392266}
\urldef\tempurl%
\url{https://doi.org/10.1145/3510457.3513044}
\showDOI{\tempurl}


\bibitem[Zimmermann et~al\mbox{.}(2019)]%
        {npm-security-threats}
\bibfield{author}{\bibinfo{person}{Markus Zimmermann},
  \bibinfo{person}{Cristian-Alexandru Staicu}, \bibinfo{person}{Cam Tenny},
  {and} \bibinfo{person}{Michael Pradel}.} \bibinfo{year}{2019}\natexlab{}.
\newblock \showarticletitle{Smallworld with High Risks: A Study of Security
  Threats in the Npm Ecosystem}. In \bibinfo{booktitle}{\emph{Proceedings of
  the 28th USENIX Conference on Security Symposium}} (Santa Clara, CA, USA)
  \emph{(\bibinfo{series}{SEC'19})}. \bibinfo{publisher}{USENIX Association},
  \bibinfo{address}{USA}, \bibinfo{pages}{995–1010}.
\newblock
\showISBNx{9781939133069}


\end{thebibliography}
